\documentclass[12pt]{article}
\usepackage{isa_rmbs}
\usepackage{amsmath}
\usepackage{subfig}

\def\BibTeX{{\rm B\kern-.05em{\sc i\kern-.025em b}\kern-.08em
   T\kern-.1667em\lower.7ex\hbox{E}\kern-.125emX}}
\usepackage{wrapfig}				
\usepackage{graphicx}

\begin{document}

\title{Microtubule Motility Analysis based on Time-Lapse Fluorescence Microscopy}
\author{\textbf{S. Masoudi}\supit{a}, \textbf{C. H.G. Wright}\supit{a}, \textbf{J. C. Gatlin}\supit{b} and \textbf{J. S. Oakey}\supit{c}  \medskip\\
        \supit{a}Electrical and Computer Engineering Department\\ 
        University of Wyoming, Laramie, WY 82071 \smallskip\\
        \supit{b}Molecular Biology Department\\ 
        University of Wyoming, Laramie, WY 82071 \smallskip\\
        \supit{c}Chemical Engineering Department\\ 
        University of Wyoming, Laramie, WY 82071 \smallskip\\}

\date{} 
\maketitle
\vspace{-5mm}

\begin{abstract}
This paper describes an investigation into part of the mechanical mechanisms underlying the formation of mitotic spindle, the cellular machinery responsible for chromosomal separation during cell division. In normal eukaryotic cells, spindles are composed of microtubule filaments that radiate outward from two centrosomes. In many transformed cells, however, centrosome number is misregulated resulting in cells with more than two centrosomes. Addressing the question of how these cells accommodate these additional structures by coalescing supernumerary centrosomes to form normal spindles will provide a powerful insight toward understanding the proliferation of cancer cells and developing new therapeutics.  The process of centrosome coalescence is thought to involve motor proteins that function to slide microtubules relative to one another. Here we use in vitro motility assays combined with fluorescence microscopy to visualize, characterize and quantify microtubule-microtubule interactions. After segmenting the microtubules, their speed and direction of movement are the extracted features to cluster their interaction type. In order to evaluate the potential of our processing algorithm, we created a simulated dataset similar to the time-lapse series. Once our procedure has been optimized using the simulated data, we will apply it to the real data. Results of our analyses will provide a quantitative description of interaction among microtubules. This is a potentially important step toward more thorough understanding of cancer. 

\end{abstract}
\section{Introduction}
Cancer is a somewhat generic word that refers to a category of all diseases having out of control cell growth as their main characteristic. This non-stop cell division leads to the formation of cell masses called tumors. Tumors can interfere with the regular mechanism of digestive, nervous, and circulatory systems or release hormones that change the body function~\cite{1,Applegate_thsis}. According to the American Cancer Society, cancer is considered to be one of the leading causes of morbidity and mortality worldwide~\cite{1}. It is the second most common cause of death in the US, which accounts for nearly $1$ out of every $4$ deaths~\cite{1}.

Studies over a century ago proposed aneuploidy (the presence of an abnormal number of chromosomes in a cell) as the original source of cancer~\cite{3}. Aneuploidy is the most common genomic instability in human cancer cells~\cite{3}. The prevalence of centrosome amplification (i.e., in terms of number) in a variety of tumor types and its direct correlation with aneuploidy, have provided a strong foundation for at least one possible pathway to aneuploidy in cancer cells~\cite{3}. 

To better understand the circumstances under which cancer cells might form, one should first know about the mitotic spindle structure, components and mechanism. Mitotic spindle is considered as a machine that segregates chromosomes into two daughter cells during a certain type of cell division called mitosis\cite {3}. Main elements of the spindle are MT polymers, whose intrinsic polarity and dynamic properties are crucial, for its proper function~\cite{3}. In most cell types, MT nucleation occurs primarily at two spindle poles called centrosomes~\cite{3,4}. Normally, the opposing nature of dual centrosomes facilitates the assembly of a bipolar spindle and ensures equal distribution of the replicated genome to each daughter cell~\cite{3}. Under the abnormal cancerous conditions, the overproduction of centrosomes leads to the formation of multi-polar spindles, lagging chromosome and chromosomal instability (CIN)~\cite{3, 33}. Many other factors, such as molecular motors and regulators of MT dynamics, are associated with organizing the spindle~\cite{3}.

 The past decade has provided a wealth of information on the molecular organs which are responsible for spindle assembly. Among this information, high-resolution views of the detailed movements and dynamics of the spindle MTs and chromosomes are of great interest~\cite{4}. A closer look into the mitotic spindle revealed that MTs are its essential puzzle components. These polymers are arranged head to tail to form a well-organized network of hollow tubes~\cite{5}. These tubes have two ends with different dynamic properties~\cite{5}. In the cytoplasm environment, the minus ends are found throughout the spindle and some are anchored to the centrosome while the more dynamic free plus ends probe to reach specific targets during a search and capture process~\cite{5}. This dynamic behavior of MTs is tightly regulated both spatially and temporally~\cite{7}. The behavior itself, and its regulation are of great importance for certain cellular functions such as cell division, signaling, polarity, shape maintenance and so on~\cite{7}. Thus, modeling MTs' dynamics and regulations is a topic of major interest for researchers. Results of studying their dynamics, movements, assembly, and how they interact with each other to form the spindle, will potentially yield important insights into cancer cell biology~\cite{8}.
 
 While significant investigations have been dedicated to this area, further study on how exactly MTs participate in this process could shed more light on new aspects of spindle assembly and, at a very basic level, the mechanistic underpinnings of cancer. Former studies have attempted to track individual MTs through \textit{in-vitro} motility assays, disregarding the manner in which MT-MT interaction influences the MT movements/dynamics. MTs must interact with each other to fulfill several cellular functions, the most critical of which is chromosome segregation during mitosis. A subset of MAPs have inherent ATPase activity and function as motors which can walk along MTs. During mitotic spindle assembly, two motors are particularly important:  the minus-end directed motor cytoplasmic dynein (dynein) and the plus-end directed kinesin motor known as Eg-5.  These motors bind to, move, and spatially organize MTs relative to one another~\cite{9,Bruno2011,Castillo2007,Chen2015}. Image analysis used in this study should therefore inform us about MT velocities at each frame, elucidating the function of MAPs and their contribution to MTs sliding, interaction, and assembly~\cite{Bruno2011,Castillo2007,Chen2015}.

For this purpose, $2$-D microscopy time-lapse images of MTs' motility and interaction were acquired. These images are obtained using time-lapse total internal reflection fluorescence (TIRF) microscopy. The MTs in these images are assembled in from Xenopus egg extracts spiked with fluorescently labeled tubulin heterodimers ~\cite{7} Currently, manual inspection of these time-lapse images is the only available method, which is laborious, time consuming, heavily biased and error-prone. Accordingly, there is a need for an automated method which works much faster and provides better accuracy based on an efficient low-cost algorithm. Image processing tools will greatly automate the collection of a wide variety of metrics related to the MT-MT interactions. Conventional image processing approaches in molecular cell biology, typically consist of two subsequent steps of segmentation and object tracking. Segmentation has been usually done based on thresholding, multiscale analysis using the wavelet transform, or model fitting while nearest-neighbor or smooth-motion criteria are basic methods for the second step of data association~\cite{smal2008particle}. Spatiotemporal-based methods and particle filtering have also been proposed respectively for MT growth analysis which demonstrate to work well~\cite{smal2008particle}. Metrics of interest in this project include whether or not MTs interact with each other while passing nearby. To accomplish this aim, two main contributions are provided in this paper. First a new model for simulating the MTs' motility has been proposed that closely represents the actual time-lapse images. Next, some standard deterministic methods have been applied to this simulated data for performance evaluation of these algorithms in a controlled way.

\indent To substitute the manual inspection which is the currently used method for detecting the potential MT-MT interaction in time-lapse images of MTs, an automatic image processing algorithm is required. Image processing tools will greatly automate the collection of a wide variety of metrics related to the MT-MT interactions. Results of studying their dynamics, movements, assembly, and how they interact with each other to form the spindle, will potentially yield important insights into cancer cell biology. 

The remainder of this paper is organized as follows. Section $2$ describes the simulated data and introduces the processing algorithm. Section $3$ provides the results generated by our processing algorithm on simulated data and finally sections $4$ and $5$ present our conclusions.

\vspace{-5mm}
\section{Methods}

\subsection{Simulated Data}
As was previously mentioned, the first step in this research project is to \textcolor{black}{create a simulated dataset which can approximately imitate the appearance and dynamic behavior of MTs that are observed in the actual microscopy time-lapse images. Similar to other cases where AI is applied to solve the biomedical problems, main reason for generating the simulated data is to compensate for the limited annotated data available for training. Besides,} the simulated data is used as input to the proposed image processing approaches which identify individual MTs and track them in sequential time-lapse image frames. The outcomes of this process will provide the basis with which to examine, evaluate and make improvements to the processing algorithms.

Similar to the provided time-lapse series, the simulated dataset is a $2$-D+t ($x$ and $y$ spatial dimensions plus time) signal constructed with a sequence of artificial images, each of which is assumed to be a sample at a certain point in time. Considering the camera's field of view to better resemble the entrance and egress of the MTs to a time-lapse image frame, the central part of a large predefined area is considered to be the main recorded frame. To reproduce an environment with different conditions existing in the real time-lapse images, the simulated data generator is coded in a flexible way so that the number and size of the time-lapse image frames, time resolution, number of participating MTs and the chance of MTs to interact are adjustable. To assemble the final time-lapse image a specific number of produced frames are attached sequentially.

Approximate simulation of the stochastic behavior of the MTs has been fulfilled through several assumptions about their geometric specifications and movement parameters. The geometric considerations mainly include the designed structure of MTs. Each MT can be defined as a wagon train containing certain number of wagons or blocks, where each block or wagon is formed by ${w\times w}$ pixels. The first block is the head wagon while the last is its tail. These wagons overlap to a certain amount that can be controlled by the code generator. Other details are the length and width of an artificial MT. Overall length is determined primarily by the number of wagons. 
For the length to be realistically random, it is determined via a pseudo-random number generator (PRNG) having a uniform probability distribution over an interval of $(\frac{L}{2},L)$. $L$ limits the maximum possible length of the MTs. Fortunately, all real-world MTs have basically the same width of $w$, specified by the number of pixels along $x$ and $y$ axes in a block. Also incorporated into the program is the real-world characteristic that length might vary over time, which is achieved through defining $dL$ (the ratio of maximum length change) and the uniform probability distribution over the interval of $(-{dL},{dL})$.

Each MT has its own random initial conditions, including the coordinates of the head block, initial direction and body shape. The initial location of the head block follows a uniform probability distribution all around the main frame. Direction only has four initial possible values: ${\{{\frac{\pi}{4}},-{\frac{\pi}{4}}, {\frac{3\pi}{4}},-{\frac{3\pi}{4}} \}}$ radians. \\
\begin{figure*}[!ht]
\centering
\includegraphics[width=0.85\textwidth]{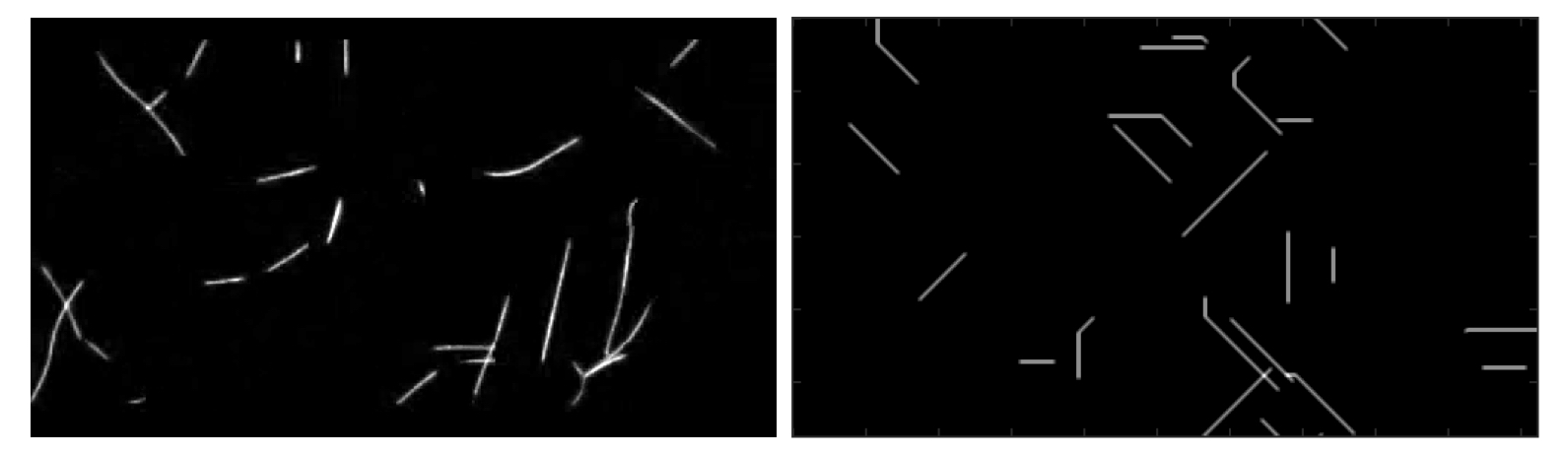}
\caption[Comparing real frame to simulated frame]{A $288 \times 512$ frame of (left) real frame obtained from sliding MTs which belong to Xenopus egg extracts, (right) simulated data.}
 \label{myfig1}
\end{figure*} 
For reference, $0$ and $\pi /2$ radians would be seen as the left to right horizontal and downward vertical directions, respectively. To obtain a random initial direction, basic values are used to create the first frame. Then, during the transition into the next few frames, stochastic moving parameters help to turn MTs' direction from a deterministic to a random nature. Disregarding these first few frames for the final time-lapse image compilation, results in random initial direction in the time-lapse image. This would save us from adding to the number of possible initial directions and employing a random initial body shape method at the same time. Using these initial conditions, coordinates of each of the MT pixels are exactly determined at the first frame.

Other aspects to the program are motility parameters that consist of its direction and speed. Because real images of the MTs do not show them to have highly ``bent" shapes, available choices of moving directions should not be more than $\pm{\frac{\pi}{4}}$. We implemented three allowed possibilities: \{straight, slightly right and slightly left\}. Using a Gaussian probability distribution with its mean value mapped to the straight direction, a random direction of movement is obtained. Change of direction is allowed in every five frames, again to improve the realism of the simulation. Speed also follows a uniform probability distribution that is limited between user-specified minimum and maximum values in units of pixels.

Finally, it should be noted that each pixel in a block that makes up that part of the MT adds 32 levels of grayscale to the zero background (the full grayscale range of the pixels is 256). Adding grayscales makes the inner parts of the MT image lighter than its boundaries and also leads to much lighter image regions at the areas over which MT-MT interactions occur. This contrast provides greater fidelity to the real-world microscopy time-lapse images.

\textcolor{black}{In our study, we generated 20 simulated sequences, each containing 379 frames of size $256\times256$ pixels. We used 4 randomly chosen sequences for the test data and the rest for the training data.}

 \subsection{Processing: segmentation}
 Image processing algorithms that are being used for this project can generally be grouped into two categories: $1)$ those that process individual frames and sequence them along the time axis later, or $2)$ those that process multiple frames at the same time to track an object (a particular MT) along sequential frames. The first method ignores any knowledge about possible correlation among the following frames, and therefore may lead to greater error. The second method, while more complicated, seems more appropriate and will be discussed in more detail. It should be noted that this processing is performed off-line, as real-time operation is not required.

The particular goal of the initial phase of this research project is to achieve a quantitative estimate of MTs' interaction when a certain number of them are presented in a given frame. To clarify the definition of interaction between MTs, one should understand that two MTs apparently crossing each other in the time-lapse image does not necessarily mean that they are interacting. In fact, they might only pass each other without any physical connection at all. The two discriminating features that help classify between MTs that only pass each other versus those that interact with each other are their direction and velocity while crossing. Two MTs just passing each other exhibit no significant change in their velocity and direction functions, while interacting MTs will show a completely different behavior. The next step is feature extraction which requires the exact segmentation of each of the crossing MTs first.

 Segmenting a single MT in a frame that includes several MTs which may be crossing each other is not a trivial task. To accomplish this, the initial step is the very common method of thresholding the image (i.e., a single frame). Having obtained a binary thresholded image, the next step is to apply the region growing algorithm to find all the neighboring non-zero gray-level pixels indicating distinct objects. This is accomplished based on a simple iterative searching algorithm for non-zero values in an 8-pixel wide neighborhood that starts from an initial pixel.
 As a result we will get some detected regions which might belong to a single MT or a crossing situation. Obviously, in case of a single MT the segmenting mission is already accomplished, while crossing regions are the target of more processing to extract single MTs out of a crossing situation. Crossing regions can be distinguished by their higher than usual gray level (due to the added gray-level per each MT's pixels, the crossing area is at least two times``brighter" than its surrounding pixels). Since this simple sequence of methods (thresholding and region growing) fails to segment individual MTs out of a crossing situation, several methods have been explored for this purpose. The remainder of this subsection describes these methods, along with their cons and pros.

 \subsubsection{Template Matching}
Template-matching is the first applied method to segment single MTs out of crossed sections. Template matching finds the spatial position of a given pattern by performing a pixel-wise comparison of the image to the given template of desired pattern [9]. However, defining an exhaustive collection of all possible crossing scenarios and picking the most similar template to properly segment passing MTs seems impractical. 
 The matching criteria can be defined in various ways: correlation function, squared error, or a simple sum of pixel by pixel multiplication. \textcolor{black}{Herein, we used 3 lengths of 5, 15, and 50 pixels along with 4 different orientations of: ${\{{\frac{\pi}{4}},{\frac{\pi}{2}}, {\frac{3\pi}{4}},\text{ and }{\pi}\}}$ to build our pool of templates. Results of applying different criteria functions on the training data are compared and the best is used to quantify the test data in the results section.}

 \subsubsection{Gradient-based Method}
 Another method was investigated which is based on the existing contrast in each object. This method uses the relatively large gradient value at the edges to produce a binary image indicating boundary pixels. An oriented iterative searching algorithm then allocates these pixels to distinct objects. This orientation is specified by gradient vector and helps select between clockwise or counterclockwise direction of search in the 8-pixel wide neighborhood. The criteria is having the least gradient change in the extracted results. After a closed path of boundary pixels, inter-pixels would be determined next. Although implementing this method is more time intensive, it can work better in special situations compared to the previously described methods. \textcolor{black}{Herein, we utilized the training data to measure the threshold value for the gradient change under which we can discard the changes in the test data.} 

\subsubsection{Morphological-based Method}

 The next method is based on using the morphological erosion of a square-shape of size $w$ (predetermined width of MTs) on the binary thresholded version of the frames. Doing so leads to binary images consist of lines or unit-width MTs called skeletons. Removing the connector pixels (the higher gray-level ones at cross section) results in some lego pieces. Different combinations of these pieces and connectors can montage different potential segmentation solutions. \textcolor{black}{Applying the correlation-based matching criteria (which is the ideal indicator based on our training data) to the last frame's results helps to resolve the uncertainty for segmentation of the object in the current frame. Although the proposed heuristic algorithm seems to work fine, its practical implementation fails to improve the results significantly. On the other hand, this method is not a fully automatic method since it necessitates having the ground truth for individual MT segmentation in at least one initial frame.}

\subsubsection{Adaptive Template Matching}

 Another solution is to take the time resolution into account so that the information in the past can be fused into our segmentation results at the current frame. \textcolor{black}{This method is similar to the classic template matching. However, transiting through the frames, we update the pool of templates by incorporating the segmentation results from the former frame. Since correlation-based matching criteria presented superior results while processing the images from the training dataset, we utilized it for segmenting individual MTs in the test data. This method also is semi-automatic as it depends on the manual results of individual MT segmentation in at least one initial frame.}

 \subsection{Processing: Data Association in a Tensor-based framework}
General implementation of this idea creates a tensor in which, $N$ horizontal $m\times m$ planes present $N$ frames containing $m$ objects. The  $i^{th}$ row and $j^{th}$ column element at each plane keeps the value of minimum distance existing between $i^{th}$ and $j^{th}$ MTs of the relevant frame and they can be assigned to either zero or one depending on the history of their behavior. Transition to a new frame adds another horizontal plane to the tensor. These horizontal planes are symmetric and have unit value on their main diagonal. Sudden appearance and disappearance of any new MT is exposed to the tensor in form of inserting new vertical planes or zeroing out the relevant row and column for the next frames. This tensor keeps a history of MTs' relative positions toward each other along time. Any of the previously presented segmentation method followed by the template matching are used to populate the tensor. Then the same process is applied to all the frames in reverse order so that the information in upcoming frames can be used to improve the results. Any dissimilarity in reverse processing indicates further investigation may be needed. Minimizing the costs of multiple MT association along the sequential frames lead us to have the optimal extracted trajectories.

\subsection{Problem Restatement}

As it was mentioned in details before, the problem to be solved in this project is detecting the MT-MT interaction in a series of time-lapsed images. This description implies a classification problem in which one attempts to categorize the located crossing-over MTs into two primary groups of MTs just passing over each other without any interaction and MTs which are physically interacted to each other through MAPs.

\paragraph{Features of Interest:}
The discriminating features that help classify between the aforementioned categories are MTs' direction and velocity while crossing. The reason for such feature selection is rooted in the fact that two MTs just passing each other have a continuous trend in their velocity and direction functions, while interacting MTs will show a completely different behavior.

\paragraph{Feature Extraction:}
The next step is feature extraction that includes deriving velocity and direction of the MTs while crossing each other. This requires the exact segmentation of each crossing MT. 

\paragraph{Tracking and Feature Extraction:}
\noindent After a comprehensive segmentation, a sequence of the sets of pixels for each MT is obtained. Every set represents a MT at a frame and includes the pixels' spatial location and their gray-level values at that frame. By following the frames in order, one can connect a MT's relevant sets to track its path. Based on this information, the head block can be identified next. In the case of a single MT, features are defined in terms of the average moving direction and speed of the head block's pixels. In crossing situations, the same features are extracted with respect to the overlapped area. Assuming a certain pixel, moving from $(x_i,y_i)$ to $(x_e,y_e)$ in a transition over two consecutive frames, its displacement vector is calculated as in Equation \ref{Eq1}. 
\begin{equation}
\overrightarrow{d}=\begin{bmatrix} (x_e-x_i)\\(y_e-y_i) \end {bmatrix}
\label{Eq1}
\end{equation}
Thus, this pixel's direction and the speed of the movement are calculated regarding Equation \ref{Eq2} expressions in which ${t_0}$ is the frame to frame time step. Regarding the definition for direction, recall that the reference $x$ and $y$ axis are respectively in downward vertical and left to right horizontal directions and the origin is located at the very upper-left corner of the image.
\begin{equation}
\theta=\arctan\frac{(x_e-x_i)}{(y_e-y_i)},\quad V=\frac{\sqrt{(x_e-x_i)^2+(y_e-y_i)^2}}{t_0}
\label{Eq2}
\end{equation}
 \section{Results}
\begin{figure}[!ht] 
 \centering
  \includegraphics[width=0.75\textwidth]{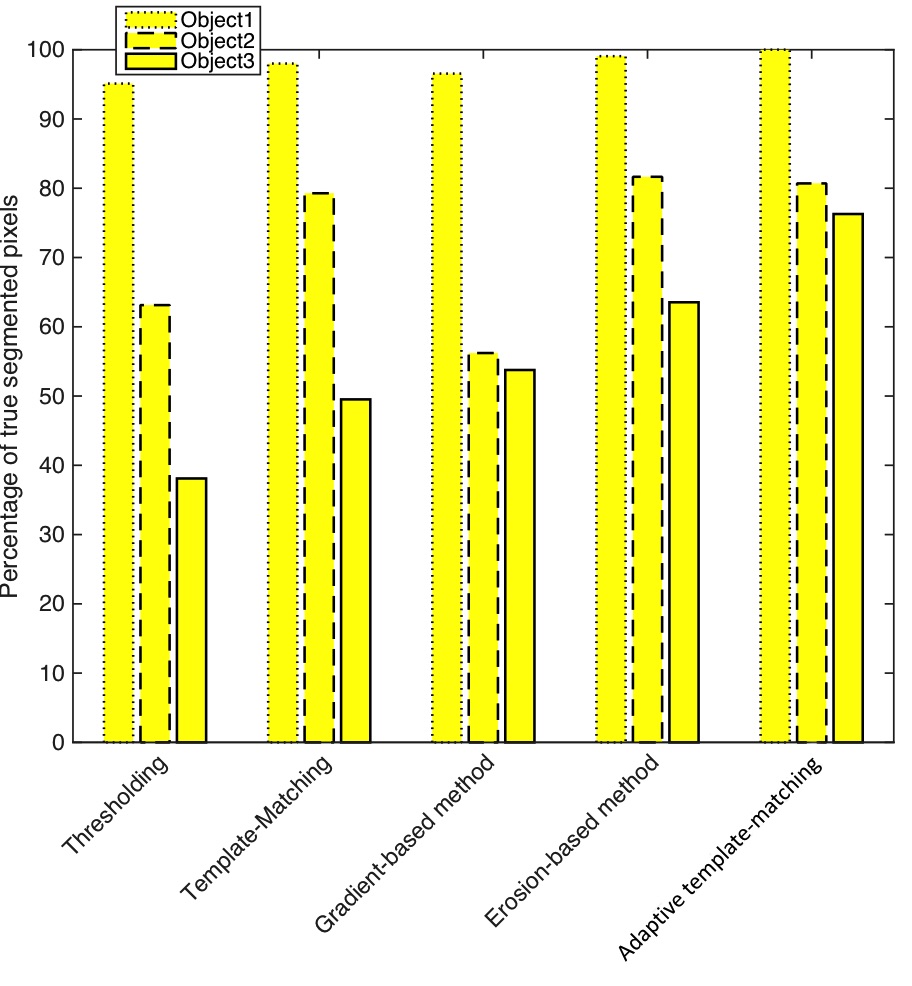} 
  \caption[Results of average correctly segmented pixels using baseline methods]{Results of segmentation in form of the average percentage of correctly segmented pixels over 20 frames.}
    \label{Plot1a}
 \end{figure}

 For segmentation, comparisons are made among all the five methods of thresholding and region growing only, template matching, gradient-based, erosion-based and the adaptive template matching, each of which followed by the tensor-based data association method. The simulated time-lapse image are provided as their input and results in terms of the pixel coordinates are obtained accordingly. \textcolor{black}{The real data contains 23 RGB, time-lapse sequences with each having a duration of $23.16\pm 12.68$ seconds. Each sequence is sampled at 16 frames per second rate with $256\times256$ pixels spatial dimension.} In order to better present the results, the percentage of the correct segmented pixels in every frame and for every object has been acquired for each method. These results are averaged over $20$ frames while tracking $3$ specific MTs and are depicted in Figure \ref{Plot1a}. According to each object's movement pattern and interactions, different results are obtained. Object $1$ is a single MT that hardly passes over other MTs during the target frames. The second object passes along another MT's length at the latter frames which brings down the segmentation results especially in case of thresholding and the gradient-based methods. 
 \begin{figure}[!ht] 
 \centering
 \includegraphics[width=0.9\textwidth]{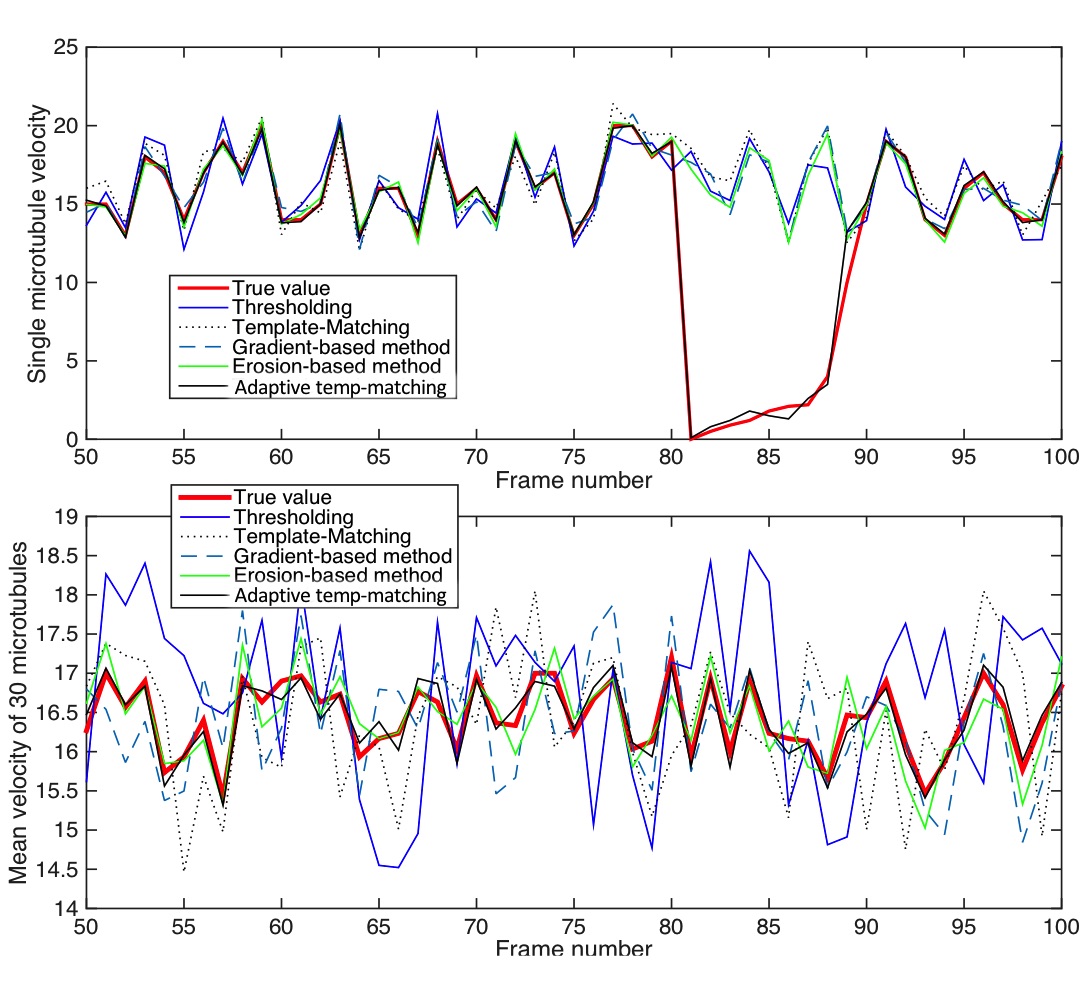} 
  \caption[Results of extracted velocity of a single MT and average of MTs using baseline methods]{Results of feature extraction in form of the extracted velocity of a single MT and the average of 30 MTs.}
    \label{Plot1b}
 \end{figure}

%
The last MT passes through another MT at the mid-frames and detached at the end. Results for this scenario look compelling. The first two methods completely fail to segment the object even after passing the crossed section in contrast to the adaptive template matching method which keeps the history of the interactions, uses the temporal information and results in the best among all.

Figure \ref{Plot1b} shows the plots of the extracted velocity using all these five methods comparing to its true value. The upper plot shows the velocity of a single MT during 50 sequential frames while the bottom one is the average velocity of all $30$ MTs along the same frames. As it can be seen in the upper plot, a change in trend of the velocity caused by the simulated interaction is happening during frame numbers $80$ to $90$ and in this special case, the only method that segmented the MT with a true estimation of velocity is the tensor-based algorithm. To have a general perspective, the average plot of the true velocities of all MTs comes with the average of velocity estimation results using the mentioned algorithms. Once again, it can be implied that the first two methods have the largest deviations from the true value of the average velocity while adaptive template matching method has the least error. However, average velocity is not a good metric for evaluation since it disguises the poor performance of an algorithm by averaging, which is not desired especially since only interacting scenarios are our actual targets. 
\vspace{-4mm}
\section{Discussion}
Although the results seem promising, there are more complicated scenarios which need to be investigated. A generalized segmentation method should capture all possible scenarios, such as situations in which a MT has interactions with more than one MT at different locations, or more than one interactions occurs at the same place, or even situations where two MTs have a long overlap along their length. The processing algorithm should consider any level of interaction at any location and discriminate between being close versus overlapping. Using temporal resolution information by keeping a history of the interactions of skeleton images and applying the correlation-based template matching showed acceptable initial results, but more accurate methods are needed for more complicated scenarios. The need for improved modeling is indicated. Regarding the inherent stochastic nature of the MTs' motility, applying probabilistic models and estimating methods may yield more reliable results.

\vspace{-4mm}
\section{Conclusions}
In order to detect microtubule-microtubule interactions several methods have been applied in this paper. Although these methods have simple implementations and provide reasonable results, their deterministic framework shows a restricted performance for a stochastic problem. Thus, applying probabilistic models may yield more reliable estimations.
Significant progress has been made to simulate and extract metrics of video microscopy of microtubules. As this research progresses, it is hoped that it will provide a better understanding of the underlying mechanics of cancer.

 \bibliographystyle{ieeetr} 

\end{document}